\documentclass[epj]{webofc}
\usepackage[varg]{txfonts}   
\wocname{EPJ Web of Conferences}
\woctitle{ICNFP 2017}
\begin{document}
\selectlanguage{english}
\title{Measurement of anisotropy of dark matter velocity distribution using directional
detection}
%
%

\author{Keiko I. Nagao\inst{1}\fnsep\thanks{\email{nagao@sci.niihama-nct.ac.jp}}\footnote[0]{This article is based on the study presented in  \cite{Nagao:2017yil}.}}

\institute{National institute of Technology, Niihama college
}

\abstract{%
Although the velocity distribution of dark matter is assumed to be generally isotropic,
some studies have found that $\sim\hspace{-0.1cm}25$\% of the distribution can have anisotropic components. 
As the directional detection of dark matter is sensitive to both the recoil energy and direction of nuclear recoil,
directional information can prove useful in measuring the distribution of dark matter. 
Using a Monte Carlo simulation based on the modeled directional detection of dark matter, we 
analyze the differences between isotropic and anisotropic distributions 
and show that the isotropic case can be rejected at a 90\% confidence level if $O(10^4)$ events can be obtained.
}
\maketitle
\section{Introduction}
\label{intro}
Dark matter accounts for about $27\%$ of the energy density of the universe \cite{Ade:2015xua},
and many direct search experiments have been conducted with the goal of detecting it. 
In ordinary direct detection, 
the recoil energy of dark matter-nucleon scattering events is measured.
 The directional
detection of dark matter, which involves the search for both the recoil energy and direction of dark matter,
has become an  active area of research \cite{Ahlen, Mayet2016}. 
Directionality is critically useful in observing the velocity distribution of dark matter.
Although velocity distribution is generally modeled as isotropic Maxwell-Boltzmann distribution,
some observations and simulations suggest the usefulness of employing an anisotropic distribution \cite{Sagittarius}-\cite{KLS}. 
In this study, the use of a directional detector to measure the anisotropy of dark matter is studied based on 
a Monte-Carlo simulation of dark matter and target scattering.
Using data produced by the simulations,
the differences between the isotropic and anisotropic cases are analyzed.

This paper is organized as follows: In Section \ref{sec:velocitydistribution}, a velocity distribution model that
includes anisotropy is introduced. Details of numerical simulation and chi squared testing for the discrimination 
between isotropic and anisotropic distributions are provided in Section \ref{sec:numericalcalculation}. 
Finally, a conclusion is presented in Section \ref{sec:conclusion}.

\section{Velocity distribution}
\label{sec:velocitydistribution}
Direct detection results can be used to derive constraints on dark matter mass and interaction using the following relation:
\begin{align}
\frac{dR}{dE_R}=\frac{\rho_\chi}{2m_\chi \mu_A^2}\int_{v_\mathrm{min}}^{v_\mathrm{max}}dv\ \frac{f(v)}{v}F^2(E_R)\frac{d\sigma(v)}{dE_R}
\label{eq:dRdE}
\end{align}
where $dR/dE_R$ is the event rate, $\rho_\chi$ is the dark matter density, $m_\chi$ is the dark matter mass, 
$\mu_A$ is the reduced mass of the dark matter and 
target nucleus, $f(v)$ is the velocity distribution of the dark matter, $F(E_R)$ is the nuclear form factor, and $\sigma(v)$ is the cross section of the dark matter.
Generally, the velocity distribution $f(v)$ is assumed to be an isotropic Maxwell-Boltzmann distribution;
however, the existence of anisotropic components is suggested by observations of the Sagittarius dwarf galaxy \cite{Sagittarius} and N-body 
simulations \cite{darkdisk}-\cite{KLS}. In this study, we adopt the velocity distribution derived in the reference \cite{LNAT}, in which not only
dark matter particles but also baryons and gasses are included:
\begin{align}
f(v_\phi) = \frac{1-r}{N(v_{0,\mathrm{iso.}})} \exp\left[-v_\phi^2/v_{0,\mathrm{iso.}}^2\right] + 
		\frac{r}{N(v_{0,\mathrm{ani.}})} \exp\left[-(v_\phi-\mu)^2/v_{0,\mathrm{ani.}}^2\right] ,
\label{eq:doublegaussian}
\end{align}
where $N(v_0)$ is a normalization factor, $v_\phi$ is the tangential velocity of the dark matter with respect to the galactic rest frame,
and the parameters $v_{0,\mathrm{iso.}} = 250$ km/s, $v_{0,\mathrm{ani.}} = 120$ km/s and $\mu=150$ km/s
are determined through the simulation. 
In Equation~(\ref{eq:doublegaussian}), the first term and the second terms correspond to isotropic and anisotropic contributions, 
respectively.
The second term proposed to anisotropy parameter $r$. In the simulation conducted in \cite{LNAT}, the value $r=0.25$
was suggested;
in this study, $r$ is generally taken as $0 \leqq r \leqq 1$; however, we primarily focus on the case  $r=0.25$.

\section{Numerical calculation}
\label{sec:numericalcalculation}

\subsection{Setup}
A directional detector can in principal obtain both recoil energy and direction of nuclear recoil.
In this paper,  the  angle between the direction of nuclear recoil and the initial direction of dark matter particle motion is
refereed to as the scattering angle $\theta$.
Scattering event information such as this can be produced using a Monte Carlo simulation, 
with the results assessed to discriminate isotropic
from anisotropic distribution cases. In this subsection, the concept and details of  such numerical calculation are described.

\subsubsection{Data sets}
Two data sets are produced by the simulation. The first, which comprises a collection of 
events number such as $O(10^8)$, is referred to as the ideal ``template data'' set.
Note that in practice the event numbers observed in direct detection are expected to be much smaller than those in the template data. 
The other data set comprises data obtained in rather realistic situations and it is referred as the ``pseudo-experimental data'' set. 
The event numbers in the ``pseudo-experimental data'' set are
determined by the event number required to discriminate isotropic and anisotropic distributions.

\subsubsection{Details}
Dark matter and target scatterings are simulated using the Monte Carlo method as follows.
\begin{itemize}
\item Targets: Two types of target are adopted. The first is fluorine (F), which is typically used in gaseous detectors for directional searches.
The second is silver (Ag), one of main target atoms used in the solid detector, i.e., nuclear emulsion detector. Both types of target are 
commonly used in directional detectors. In this study, F is assumed to be relatively light target, while Ag is assumed to be a relatively heavy
target.
\item Dark matter mass: For simplicity, the mass relation $m_\chi=3m_A$ is assumed, where $m_A$ is the mass of the target nucleus.
\item Scattering: Only elastic scattering is considered in the simulation.
\end{itemize}

\subsubsection{Two cases depending on detector resolution}
Several cases involving different detector resolutions can be identified. 
The the first case, the detector has a high angular resolution but a low energy resolution. 
In this case, an energy histogram is suitable for practical use, as discussed in Subsection \ref{subsec:numericalresult}.
In the second case, the energy resolution is high but the angular resolution is low,
representing a situation similar to the case of ordinary direct detection; therefore, we will skip the case. 
The third case is an  ideal case in which both the 
energy and angular resolutions are high, enabling an energy-angular distribution to be directly obtained. 
The case is examined in Subsection \ref{subsec:numericalresult2}.

\subsection{Numerical result of angular histogram}
\label{subsec:numericalresult}
\begin{figure}[t!]
\centering
\includegraphics[width=14cm,clip]{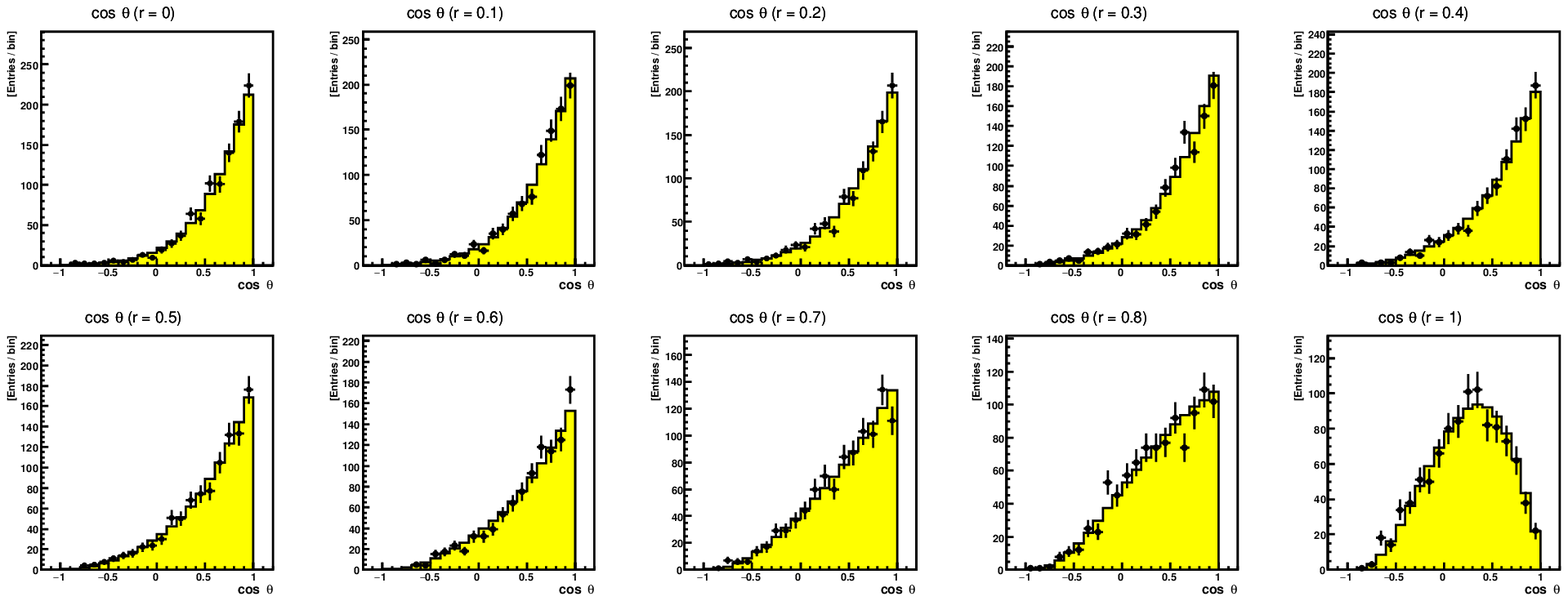}
\caption{Angular histogram for case using F as target. The yellow histogram and black points are, respectively, 
ideal template and pseudo-experimental data. 
The pseudo-experimental data event number is $10^3$. 
The energy threshold is assumed to be $20$ keV.}
\label{fig:F_1e3_20keV_CosTheta_Overlay}    
\includegraphics[width=14cm,clip]{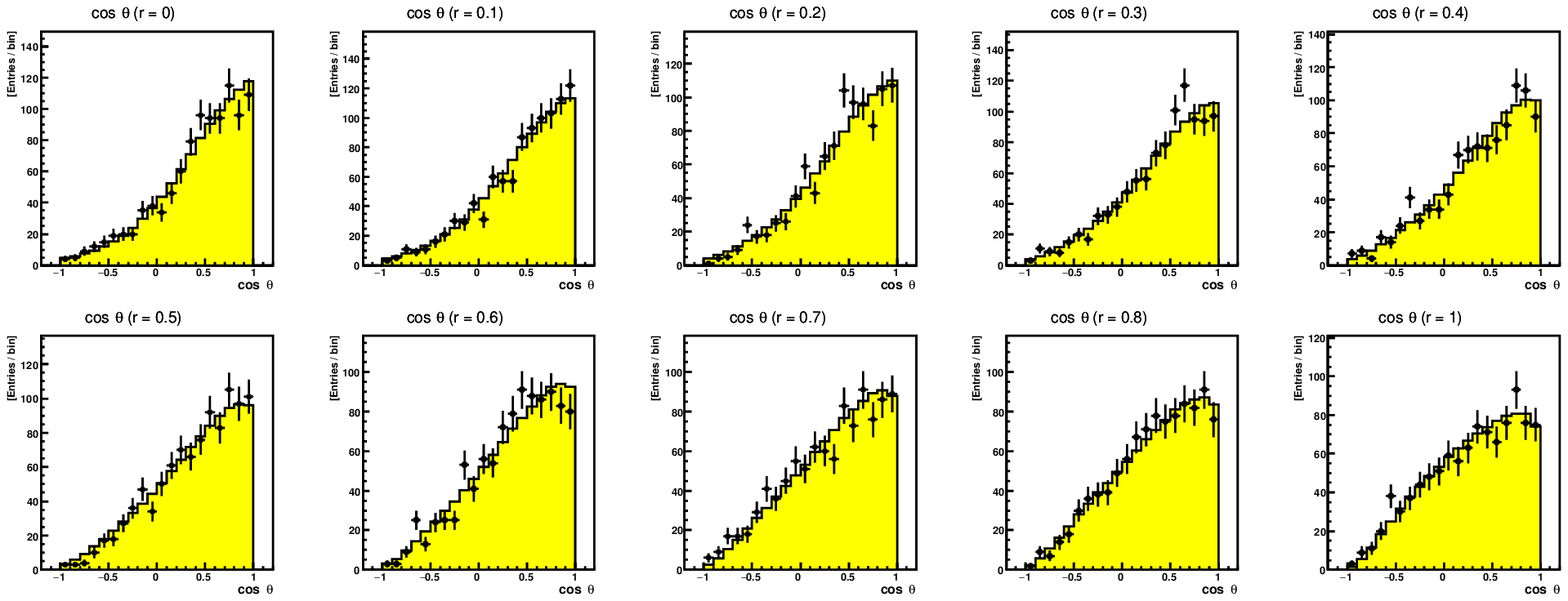}
\caption{Angular histogram for case in which Ag is used as a target. The yellow histogram and black points represent, 
respectively, ideal template and pseudo-experimental data. The pseudo-experimental data event number is $10^3$. 
The energy threshold is assumed to be $50$ keV.}
\label{fig:Ag_1e3_50keV_CosTheta_Overlay}    
\end{figure}

Even if only scattering angle data are obtained, an angular histogram can be produced. 
Figure \ref{fig:F_1e3_20keV_CosTheta_Overlay}  and Figure \ref{fig:Ag_1e3_50keV_CosTheta_Overlay} show  
angular histograms for F and Ag targets, respectively. 
In both of the figures, the event number is set at $10^3$.
The yellow histograms and black points with error bars 
correspond, respectively, to template data and pseudo-experimental data. 
At $r=0$, the peak of the histograms is 
at $\cos\theta \sim 1$. As the anisotropy grows, some of the events move to the region $\cos\theta \ll 1$;
however, there does not seem to be a significant difference 
between results at $r=0$ and those at $r=0.2$--$0.3$.

\subsection{Numerical result of energy-angular distribution}
\label{subsec:numericalresult2}
\begin{figure}[t!]
\centering
\includegraphics[width=6cm,clip]{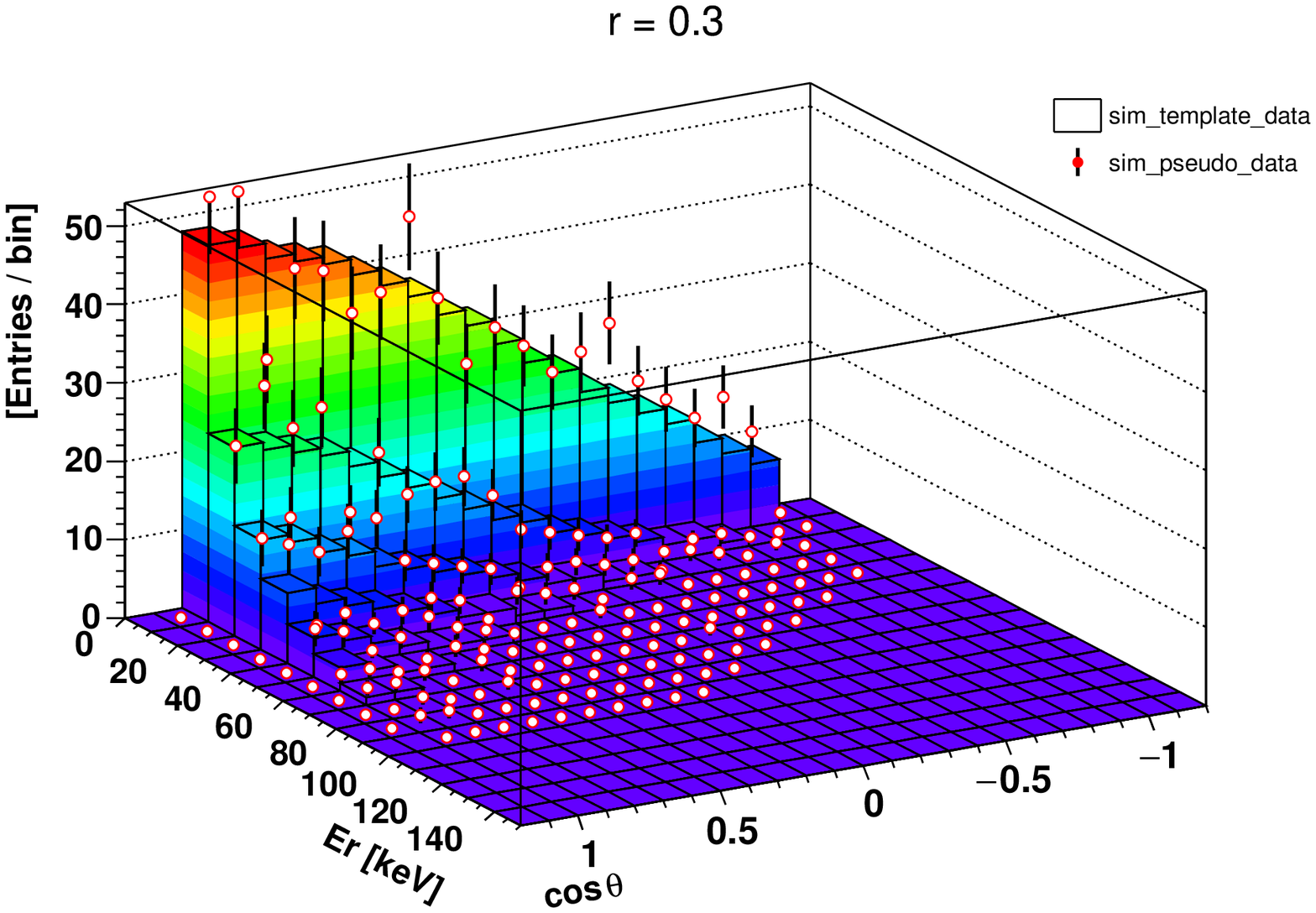}
\caption{Example of three-dimensional energy-angular distribution. Colored bars and white dots correspond, respectively, to template data and pseudo-experimental data. 
In the figure, the $E_R-\cos\theta$ plane is divided into small bins,
each of which has a calculated 
chi square value for chi squared testing.}
\label{fig:F_r03_1e3_Er_vs_CosTheta_Overlay}  
\end{figure}

\begin{figure}[h!]
\centering
\includegraphics[width=14cm,clip]{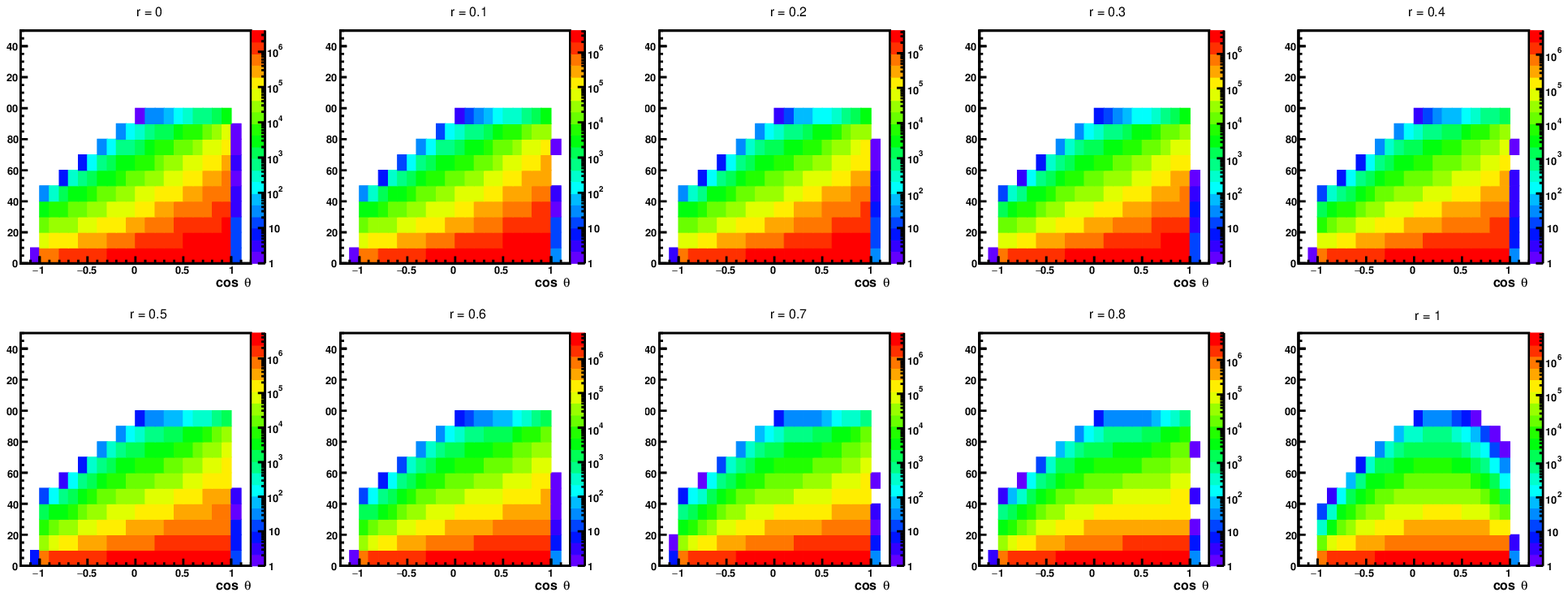}
\caption{Energy--angular distribution with target F. 
The event number of the pseudo-experimental data is $6\times 10^3$.}
\label{fig:Log_F_Er_vs_CosTheta_r}    
\centering
\includegraphics[width=14cm,clip]{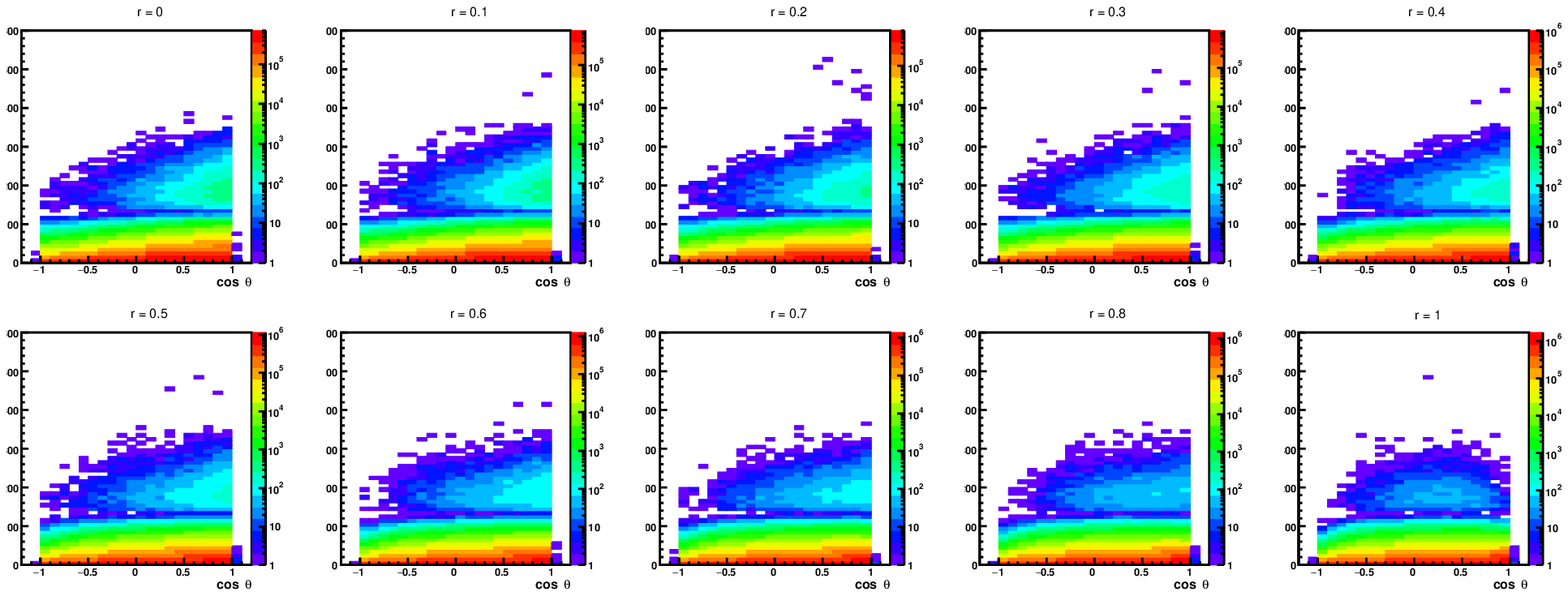}
\caption{Energy--angular distribution with target Ag. 
The event number of the pseudo-experimental data is $6\times 10^4$.}
\label{fig:Log_Ag_Er_vs_CosTheta_r}
\end{figure}

\begin{figure}[h!]
\centering
\includegraphics[width=14cm,clip]{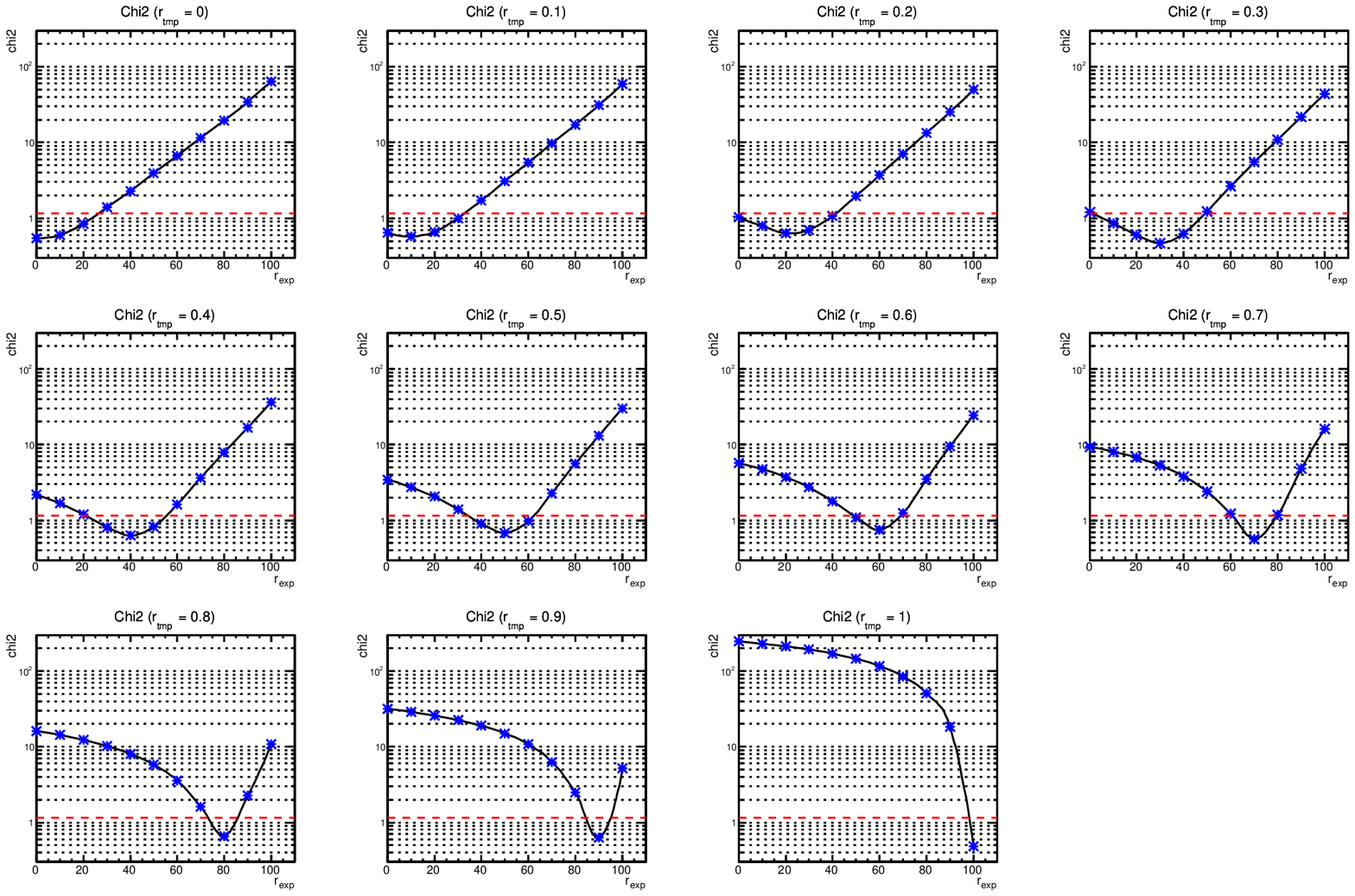}
\caption{Result of chi squared testing between template data and pseudo-experimental data for case with target F. 
The event number of the pseudo-experimental data is $6\times 10^3$. The energy threshold is assumed to be $20$ keV.}
\label{fig:F_20keV_6e3_Chi2}    
\centering
\includegraphics[width=14cm,clip]{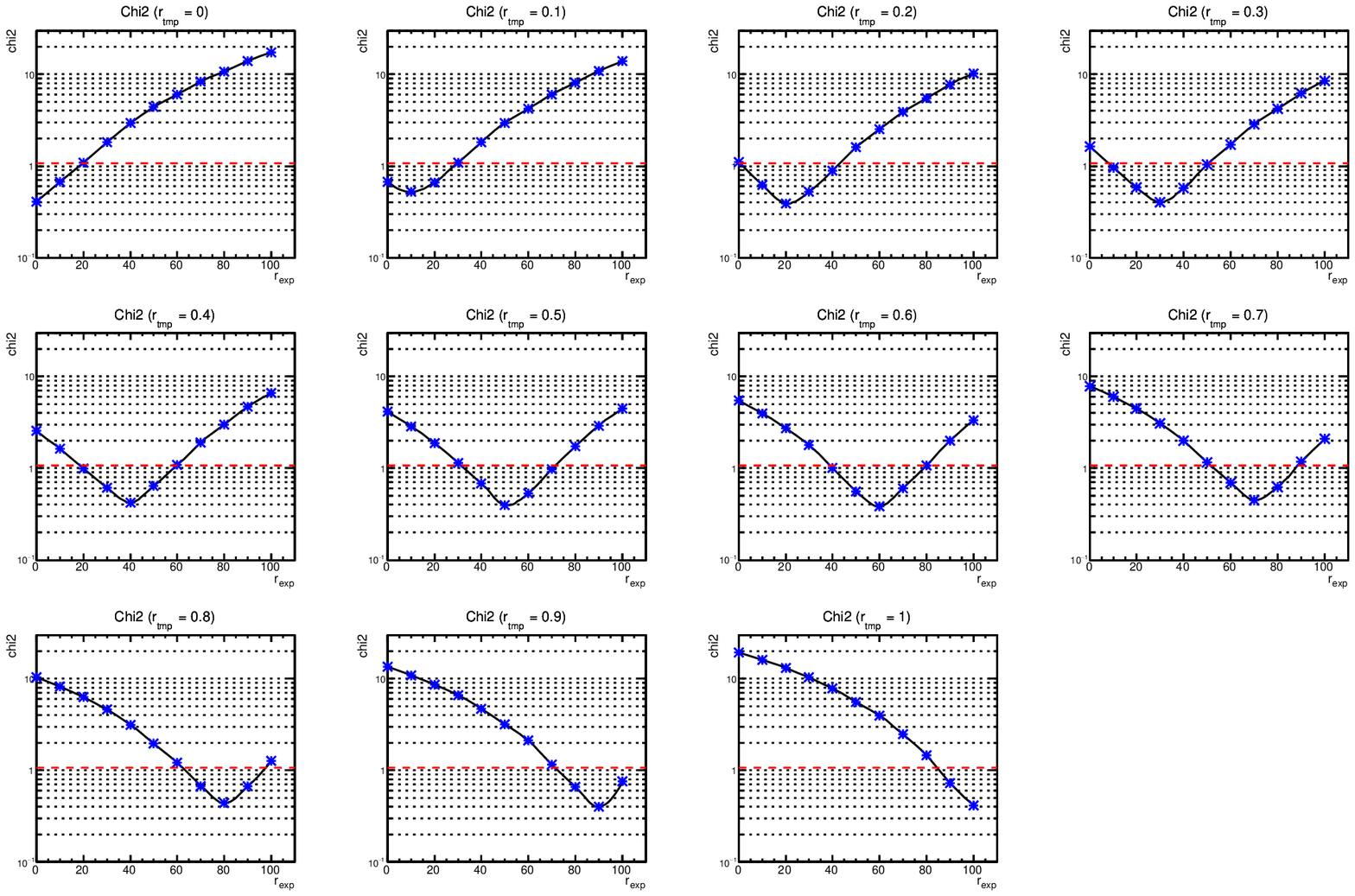}
\caption{Result of chi squared testing between template data and pseudo-experimental data with target Ag. 
The event number of the pseudo-experimental data is $6\times 10^4$. 
The energy threshold is assumed to be $50$ keV.}
\label{fig:Ag_50keV_6e4_Chi2}
\end{figure}

In case in which both the recoil energy and the scattering angle can be obtained, an energy-angular distribution can be produced.
Figure \ref{fig:F_r03_1e3_Er_vs_CosTheta_Overlay} shows an energy-angular histogram for target F and $r=0.3$  
as an example of such a distribution. To clarify the anisotropy dependence of the distribution, the $E_r-\cos\theta$ plane 
of the distribution is divided into small bins, each of which has a chi square value clculated. 
Figures \ref{fig:Log_F_Er_vs_CosTheta_r} and Figure \ref{fig:Log_Ag_Er_vs_CosTheta_r} show all of the energy-angular distributions
for each anisotropy parameter, $r$, 
and correspond, respectively, to cases in which
the targets are F and Ag and the event numbers are $6\times 10^4$ and $6\times 10^5$.

Figures \ref{fig:F_20keV_6e3_Chi2} and \ref{fig:Ag_50keV_6e4_Chi2} show the results of the chi square tests
for targets F  and Ag, respectively,  with respective set energy thresholds of $20$  and $50$ keV. 
As a reference, the same distributions for $E_R=0$~keV are shown in Appendix A.
In the figures, the red dashed lines represent the $90$\% confidence level (C.L.); i.e.,
the regions above these lines are excluded at $90$\% C.L. 
Thus, for the cases in which the anisotropy of the template data $r_{\mathrm{tmp}}=0.2$--$0.3$, 
the isotropic case $r_\mathrm{exp.}=0$ can be excluded at a 90\% C.L. The required event numbers for F and Ag targets are  
$6\times 10^4$ and $6\times 10^5$, respectively, indicating that more events are needed for discrimination in 
the heavy target case. This occurs because the effect of the nuclear form factor in Equation (\ref{eq:dRdE})
 is more significant in heavy target cases and results in strong distortion of the shape of the distribution, corresponding to more frequent rejection.

\section{Conclusion}
\label{sec:conclusion}
Although dark matter velocity is generally assumed to have an isotropic distribution, 
it can include anisotropic components. 
In this study, discrimination of this anisotropy using dark matter directional detection was assessed. 
Under the assumption of  typical light (F) and heavy (Ag) targets, 
template data including ($O^8$) event numbers and pseudo-experimental data were produced via 
Monte-Carlo simulation of dark matter--target scattering.
Based on the template data, 
the isotropic case could be excluded at a 90\% C.L. 
if an anisotropic distribution was realized. 
An event number of $O(10^{4})$ was found to be necessary for discrimination.


\section*{Appendix A}
\begin{figure}[h!]
\centering
\includegraphics[width=14cm,clip]{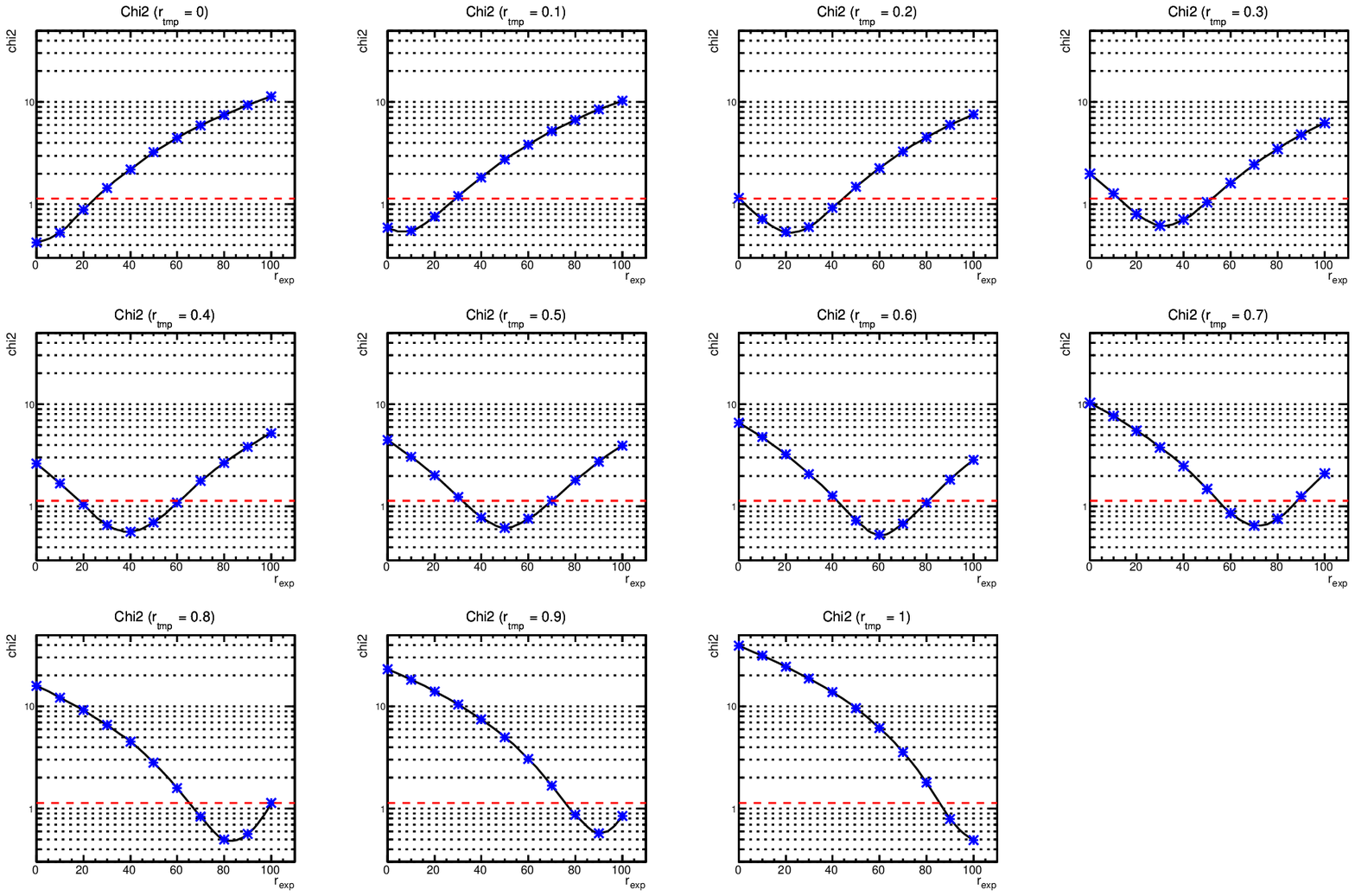}
\caption{Result of chi squared test between template data and pseudo-experimental data for case of target F. 
Event number of pseudo-experimental data is $6\times 10^3$. Energy threshold is supposed to be $0$ keV.}
\label{fig:F_0keV_6e3_Chi2}    
\includegraphics[width=14cm,clip]{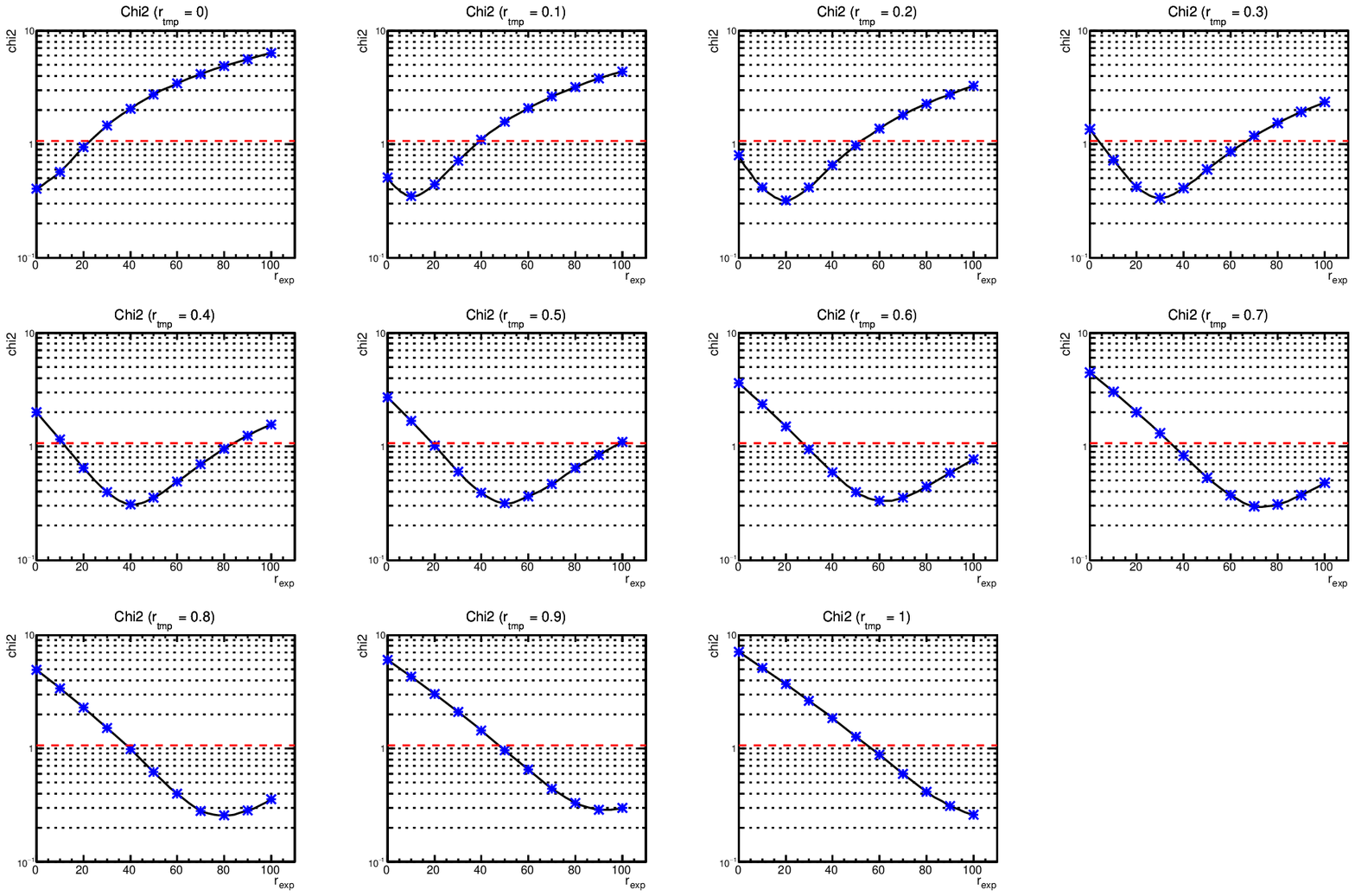}
\caption{Result of chi squared test between template data and pseudo-experimental data for case of target Ag. 
Event number of pseudo-experimental data is $6\times 10^4$. Energy threshold is supposed to be $0$ keV.}
\label{fig:Ag_0keV_6e4_Chi2}    
\end{figure}

In Figure \ref{fig:F_0keV_6e3_Chi2} and Figure \ref{fig:Ag_0keV_6e4_Chi2}, results of 
chi squared test for $E_R=0$ keV are shown as reference of Figure \ref{fig:F_20keV_6e3_Chi2}
and \ref{fig:Ag_50keV_6e4_Chi2}, respectively.

\end{document}